# Cs-induced charge transfer on (2x4)-GaAs(001) studied by photoemission

O. E. Tereshchenko,[a]* D. Paget,[b] P. Chiaradia,[c] F. Wiame,[d] A. Taleb-Ibrahimi[e]

[a] Institute of Semiconductor Physics, Novosibirsk State University, 630090 Novosibirsk, Russia,

[b] Laboratoire de Physique de la Matière Condensée, Ecole Polytechnique-CNRS, 91128 Palaiseau cedex, France,

[c] Dipartimento di Fisica, CNISM Unit and NAST, Universita di Roma Tor Vergata, 00133 Roma, Italy

[d] Laboratoire de Physico-Chimie des Surfaces, Ecole Nationale Supérieure de Chimie de Paris, 75231 Paris cedex 05, France

[e] Synchrotron SOLEIL, Saint Aubin , BP 48, 91192 Gif sur Yvette Cedex, France.

Abstract

Cesium adsorption on (2×4)-GaAs(001) was studied by photoemission and low - energy electron diffraction. The different Cs-induced changes of As 3d and Ga 3d core level spectra show that charge transfer is almost complete for Ga surface sites, but is negligible to surface As at a coverage $\Theta_{Cs} < 0.3$ ML. The situation becomes opposite for $\Theta_{Cs} > 0.3$ ML, at which transfer occurs to As but no longer to Ga. Charge transfer to As atoms leads to surface disordering and destabilization and induces surface conversion from As-rich to Ga-rich (4x2)-GaAs(001) surface after annealing at a reduced temperature of 450°C.

Adsorption of alkali metals on semiconductor surfaces has been investigated in most detail for silicon[1,2] and for the (110) cleavage face of GaAs,[3-5] for which the nature of chemical bonding, the amount of charge transfer, and the origin of the surface dipole have been essentially clarified. On the other hand, despite of technological applications, the fundamental aspects of adsorption on GaAs(001), including the difference in the adsorption mechanisms at the empty gallium-related and at the occupied arsenic-related dangling bonds is still far from being known. For the (2×4) As-rich surface, the early stages of Cs adsorption, characterized using both X-ray diffraction and theoretical calculations,[6] give rise to adsorption near both As dimers and Ga dangling bonds with a significant charge transfer only in the latter case.

Here, we report investigation of the microscopic nature of the bonding between Cs and (2x4)-GaAs(001) using photoemission spectroscopy. The investigations were performed at the SU3 beamline of the SuperAco storage ring (Orsay, France), in a UHV chamber with a base pressure in the $10^{-11}$ mbar range. The photoemission spectra were obtained using a hemispherical electron analyzer, at an overall energy resolution of the order of 0.15 eV at photon energy of 100 eV. Clean reconstructed surfaces of epitaxial p-GaAs(001) were obtained by using a treatment by HCl in isopropyl alcohol under dry nitrogen atmosphere, followed by introduction into ultra-high vacuum (UHV) and annealing to 450°C.[7] Low energy electron diffraction (LEED) investigations and measurements of the work function changes were performed in an independent setup by a retarding field method using a LEED gun.[8] Several successive exposures to cesium (20, 30, and 50 minutes) were performed at RT using thoroughly outgassed dispensers in a vacuum not exceeding $10^{-10}$ mbar. Using a previous Auger investigation,[9] we estimate that a deposition time of 50 minutes corresponds to a cesium coverage $\Theta_{Cs}$ of about 50% of the saturation concentration, the latter concentration being taken as one monolayer (ML). The Cs 4$d$ core level spectra (CLs), not shown here, are similar to those obtained for low coverage at low temperature.[10] Their shapes, consisting of one broad doublet, do not significantly change with coverage, and their intensities stay approximately proportional to the Cs exposure.

Shown in Curve a and Curve b of the left panel of Fig. 1 are the As 3$d$ CLs before and after Cs adsorption, respectively. The Cs-induced effect, found from the difference a-b shown in Curve d, corresponds to the disappearance of a doublet-like contribution lying at higher binding energy (BE). Also shown in the right panel of Fig. 1, are the corresponding Ga 3$d$ CLs, as well as in Curve i the difference showing the effect of Cs adsorption. This curve shows the decrease of a component at higher binding energy. Also shown in Curve c and Curve h are the CLs of the clean Ga-rich surface, obtained by annealing the clean As-rich one to 550°C, as well as the annealing-induced changes of the CLs (obtained from the differences between a and c and between f and h, respectively). Note that the annealing-induced change of the As 3$d$ CLs (Curve e) is remarkably identical to the Cs-induced one.(Curve d)

Shown in Fig. 2 are the decompositions of the As 3$d$ and Ga 3$d$ CLs into bulk contributions and surface components (S for Arsenic and S' for Gallium), using the shapes and chemical shifts of surface components identified in Fig. 1. For Ga 3$d$ (As 3$d$) CLs the Lorentzian width was held constant at 0.16 eV (0.18 eV) and the Gaussian width was allowed to vary as a fitting parameter in the 0.50-0.60 eV (0.55-0.70 eV) range, while standard values of the spin-orbit splitting SO = 0.44 eV (0.69 eV) and of the branching ratio in the range R=1.50-1.55 were used.[11] For As 3$d$ at $\Theta_{Cs}$<0.3ML , we have used two surface components. There is first a contribution $S_1$, shifted by 0.62 eV to the higher BE side. This contribution is known to be caused by excess arsenic under the form of As dimers which are bonded to As of the second layer.[12,13] Also present is a contribution at a chemical shift of -0.44 eV, labeled $S_2$, and already observed before at similar chemical shift.[11,14-17] The most reasonable interpretation is in terms of dimerized As atoms because the positive sign of the chemical shift is in agreement with the excess electronic charge due to the occupied dangling bond of the top As atoms which are bonded to Ga of the underlying layer.[12] For the Ga $3d$ CLs we used a surface component labeled $S'_1$, at a chemical shift of 0.49 eV, reminiscent of the signal in Curve i of Fig. 1 and also reported before.[14] We attribute it to second layer Ga atoms for which the empty dangling bonds generate a shift to lower energy with respect to the bulk configuration.

Summarized in Fig. 3 are the Cs-induced changes of the surface components, with respect to the intensity of the bulk contribution. Also shown for comparison are the changes of the work function under Cs deposition. Cs-induced *As desorption* is revealed by disappearance of the $S_1$ component, due to excess arsenic. It is known that adsorption of an individual Cs atom at GaAs releases an energy of the order of 3 eV at the initial stage of adsorption and about 1.5 eV at about half of saturation coverage.[18] Such energy can readily induce desorption of the weakly-bonded excess arsenic atoms. The decrease of the As surface concentration is also confirmed by a decrease of the As/Ga ratio by about 15% at the initial adsorption stage. Other changes of surface components are different for As $3d$ and Ga $3d$ and reveal two distinct regimes, below and above $\Theta_{Cs}=0.3$ ML (30 min Cs deposition), respectively.

Up to 0.3 ML, the $S_2$ component of the As $3d$ CLs stays constant while, for Ga $3d$, the $S'_1$ component decreases by almost 30%. The LEED pattern shows an increase of the background level, revealing some surface disordering but, at this coverage, the (2×4) structure is still visible. The results of ab-initio calculations[6,19] for the initial stage of adsorption (one Cs atom per unit cell) show that the Cs atoms can adsorb at both As-related and Ga-related sites such as the dimer site D and the $T_3$ trench site near surface arsenic and the site $T'_2$ near the empty dangling bond of second layer gallium atoms.[20] Cs adsorption only weakly modifies the surface geometry and the positions of surface atoms, so that changes of surface components can only be due to Cs-induced modifications of the electronic environment of the relevant atoms. The constant value of the As dimer-related $S_2$ component shows that charge transfer of Cs adsorbed at the As dimer site is negligible. In contrast, the decrease of the Ga-related $S'_1$ component reveals the change of the electronic environment of second layer Ga atoms due to a transfer of charge from Cs atoms adsorbed at the nearby $T'_2$ site. The latter results are in complete agreement with the theoretical predictions, according to which the charge transfer is negligible for the dimer site and of the order of $\delta n_{Ga}$ ~0.7 electron for the $T'_2$ one.[6] Such transfer, also predicted by simple tight-binding calculations including the empty Ga dangling bond,[10] should lead to a decrease of the magnitude of the chemical shift of $S'_1$, given at low-coverage by $\delta\Delta E = \Delta E_{clean} + \delta n_{Ga} \frac{e^2}{\varepsilon_b}\left(\frac{1}{r_{Ga}} - \frac{1}{d_{Cs-Ga}}\right)$, where $\Delta E_{clean} = -0.49\ eV$ is the chemical shift of $S'_1$ on the clean surface,[4] $r_{Ga} = 0.126\ nm$ is the gallium covalent radius and $d_{Cs-Ga} = 0.384\ nm$ is the bond length.[6] One finds that the two terms of the latter equation cancel each other so that $\delta\Delta E \approx 0$. As a result, the $S'_1$ component should shift towards lower BE up to approximately the position of the bulk contribution, which explains that Cs adsorption induces a decrease of the $S'_1$ component, without inducing the appearance of a new component. From the 30% decrease of the $S'_1$ component obtained for $\Theta_{Cs}$ =0.3 ML, we conclude that, in average, 1.3 of the four $T'_2$ sites of the unit cell are occupied by Cs atoms. Such fact is in qualitative agreement with the estimated coverage of 2 Cs atoms per unit cell at this coverage[21] and with the joint population of the Ga dangling bond sites together with the dimer sites.[6]

For $\Theta_{Cs}$>0.3 ML, three main features are outlined: i) the (2×4) LEED pattern completely vanishes (In contrast, on the Ga-rich surface, the (4×2) pattern is observed up to 0.7 ML of Cs[22]); ii) the work function changes saturate; iii) unlike the low coverage behavior, the Ga 3$d$ CLs is unmodified, while a new component labeled $S_3$ appears in the As 3$d$ CLs at a chemical shift of 0.78 eV. Quantitative interpretation of these effects requires ab-initio calculations considering several Cs atoms per unit cell and is beyond the scope of the present work. Qualitatively, the work function saturation is known to be caused by the increased interaction between Cs atoms, producing a long-range depolarizing field. The constant value of the $S'_1$ component for $\Theta_{Cs}>0.3$ ML is due to the increased interaction between surface dipoles of nearby Cs atoms, preventing further adsorption at the $T'_2$ site. The depolarization effect implies a modification of the surface electronic states induced by a condensation in the adlayer[22] and, as verified by He*-scattering,[23] charge redistribution from the top atomic layers of the substrate back to the adlayer itself. The appearance of the $S_3$ component can thus be interpreted as due to electron redistribution between second layer Ga atoms and top As atoms. The charge transfer to surface arsenic possibly occurs to the empty antibonding levels of As dimers, shown to be at a relatively low energy both by ab-initio calculations at low coverage[6] and tight-binding calculations at the clean surface.[24] Such transfer leads to a decrease of the dimer bonding energy, surface destabilization and disordering, in agreement with the observed disappearance of the (2×4) LEED pattern. The increase of the sum $S_2+S_3$ of the As-dimer-related components with respect to the bulk contribution can be explained by Cs-induced screening effects. Since As dimer sites are supposed to be occupied at earlier stages of adsorption[6] so that for $\Theta_{Cs}>0.3$ ML the Cs-induced screening mostly concern the bulk contribution and not the $S_2+S_3$ one.

In agreement with the latter results, it is now shown that the Cs-induced redistribution of the electron density also weakens the As-Ga backbonds. Cs-induced backbond weakening, already observed in Si-H,[25] is found here to occur for $\Theta_{Cs}>0.3$ML from the Cs-induced reduction of the temperature of the transition to the Ga-stabilized surface. Shown in the bottom panel of Fig. 2 are the LEED pattern and CLs obtained at RT after annealing the cesiated surface ($\Theta_{Cs}=0.5$ ML) to 450°C that is, to a temperature lower *by 100°C* than the one necessary for the clean surface. In agreement with a previous report,[26] such anneal induces a conversion to the *Ga-stabilized* surface, as seen from the clear c(8×2) LEED pattern. The dominant CLs change lies in the appearance in the Ga 3*d* CLs of a new component labeled S'$_2$ at a chemical shift of 0.43 eV. The latter component, characteristic of the Ga-rich surface,[14, 27] can be attributed to dimerized gallium atoms in the second layer of the ς model of the (4×2) unit cell.[28]

In summary, analysis of the Cs-induced modifications of the As 3*d* and Ga 3*d* CLS of (2×4)-GaAs(001) reveals : i) Cs-induced desorption of excess As, ii) For $\Theta_{Cs} < 0.3$ ML, almost complete charge transfer to the second layer Ga sites, iii) For $\Theta_{Cs} > 0.3$ ML, charge transfer to surface arsenic. Such transfer induces charge redistribution between top As dimers and underlying Ga atoms and leads to weakening of As backbonds and to a substantial reduction (by ~100°C) in the temperature of transition from the As-stabilized to Ga-stabilized surface.

Acknowledgement

The present work was partly supported by the Russian Foundation for Basic Research (grant № 09-02-01045).

Fig. 1: The left panel shows the As 3*d* CL spectrum of the clean (2x4) As-rich surface (a), after 0.3 ML Cs deposition (b), and after annealing the clean As-rich surface to 550°C, shown to reveal the clean Ga-rich surface (c). The difference spectra a-b and a-c are almost identical, which demonstrates that Cs adsorption induces As desorption from the surface. The right panel shows similar spectra and differences for the Ga 3*d* CL signal. All spectra are taken at RT.

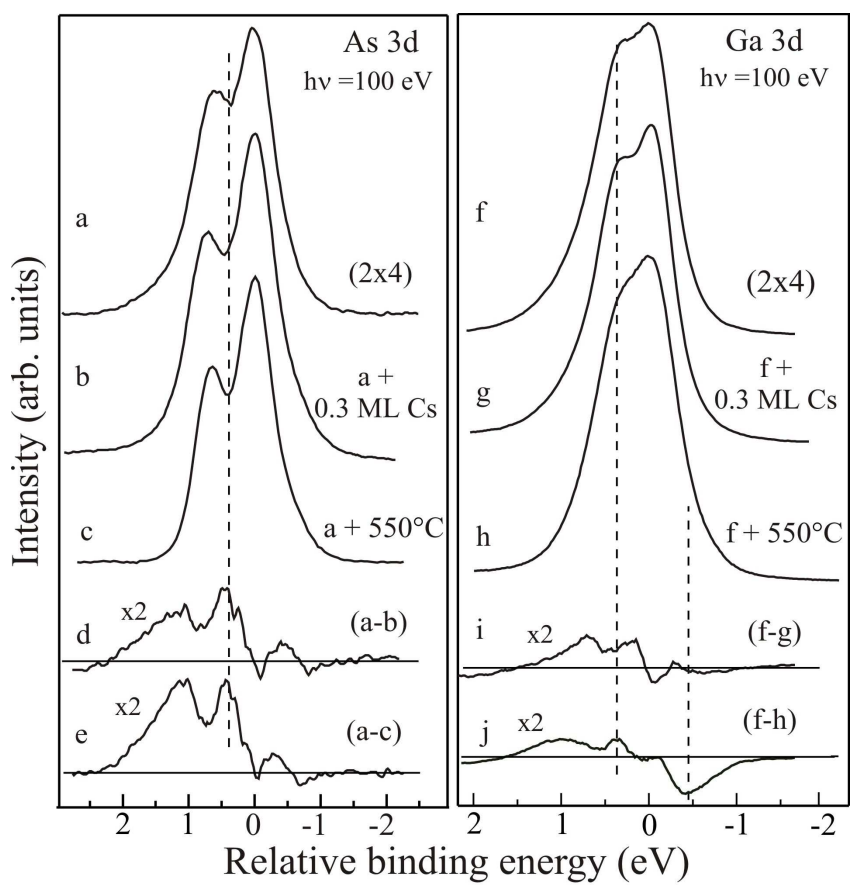

Fig.1 to the manuscript "Cs-induced..."
by Tereshchenko et al.

Fig. 2: Decomposition of As 3*d* (left panel) and Ga 3*d* (right panel) using components identified from the differences in Fig. 1. The spectra are taken for the clean surface (a), after successive Cs exposures, of total coverage 0.2 ML (b), 0.3 ML (c), 0.5 ML (d), and after subsequent anneal to 450°C (e). Also shown are corresponding LEED patterns.

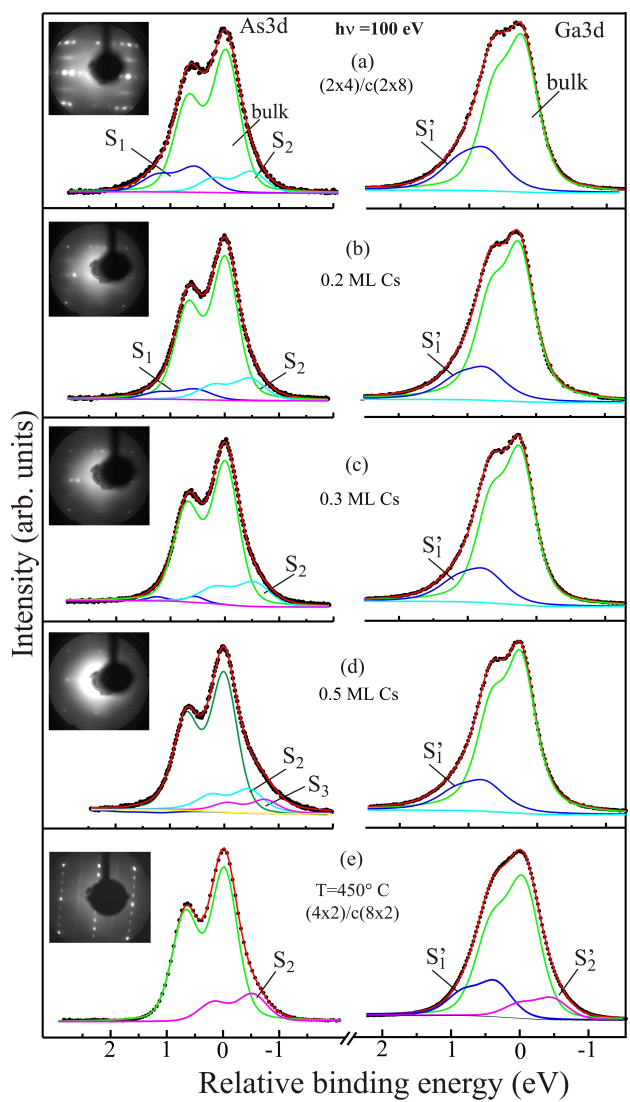

Fig.2 to the manuscript "Cs-induced..."
by Tereshchenko et al.

1/09/09         13

Fig. 3 : Variation of the work function and the intensities of the surface components of the As 3*d* and Ga 3*d* CLs.

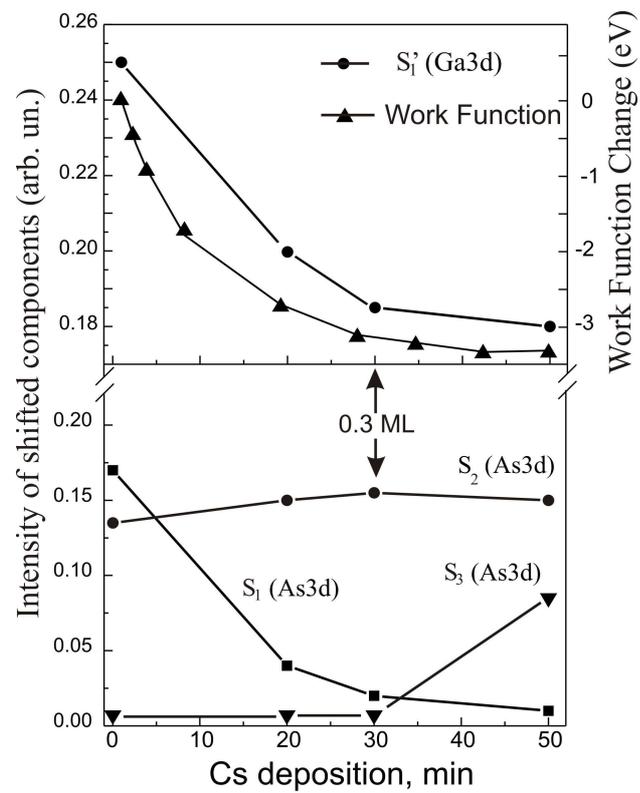

Fig.3 to the manuscript "Cs-induced..."
by Tereshchenko et al.